\documentclass[pra,twocolumn, floatfix,aps,nofootinbib]{revtex4-1}
\usepackage{amsmath,verbatim,enumerate}
\usepackage{bbm,comment}
\usepackage{amssymb}
\usepackage{graphicx}
\usepackage{times}
\usepackage{color}
\usepackage[amsmath,hyperref,standard,thmmarks]{ntheorem}
\usepackage{times}
\usepackage[normalem]{ulem}
\usepackage{etex}
\renewtheorem{theorem}{Theorem}
\renewtheorem{proposition}{Proposition}

\renewtheorem{lemma}[theorem]{Lemma}
\renewtheorem{corollary}[theorem]{Corollary}
\renewtheorem{definition}[theorem]{Definition}

\renewtheorem{example}[theorem]{Example}
\renewtheorem{remark}[theorem]{Remark}
\newtheorem{assumption}[theorem]{Assumption}

\usepackage{hyperref}

\definecolor{henrik}{rgb}{1,.0,.4}
\definecolor{cred}{RGB}{179,28,28}

\newcommand{\ii}{{\rm i}}
\newcommand{\e}{{\rm e}}
\newcommand{\ZZ}{\mathbb{Z}}

\newcommand{\NN}{\mathbb{N}}

\newcommand{\mc}[1]{\mathcal{#1}}
\newcommand{\bra}[1]{\langle #1|}
\newcommand{\ket}[1]{|#1\rangle}

\newcommand{\proj}[1]{\vert #1\rangle\!\langle#1 \vert}
\newcommand{\rmd}{\mathrm{d}}

\newcommand{\DelA}[1]{\Delta A(#1)_\Psi}

\newcommand{\norm}[1]{\left\Vert #1 \right\Vert}
\def\one{\mathbbm{1}}

\newcommand{\Tr}{\operatorname{tr}}
\newcommand{\tr}{\Tr}

\newcommand{\fu}{Dahlem Center for Complex Quantum Systems, Freie Universit{\"a}t Berlin, 14195 Berlin, Germany}

\begin{document}
\title{Towards local equilibration in closed interacting quantum many-body systems}

\author{H. Wilming, M. Goihl, C. Krumnow, J. Eisert}
\affiliation{\fu}

\begin{abstract}

One of the main questions of research on quantum many-body systems following
unitary out of equilibrium dynamics is to find out how local expectation values
equilibrate in time.
For non-interacting models, this question is rather well understood.
However, the best known bounds for general quantum systems are vastly crude,
scaling unfavorable with the system size.
Nevertheless, empirical and numerical evidence suggests that for generic interacting many-body systems,
generic local observables, and sufficiently well-behaved states, the equilibration time does not depend strongly on the system size, but only the precision with which this occurs does.
In this discussion paper, we aim at giving very simple and plausible arguments for why this happens.
While our discussion does not yield rigorous results about equilibration time
scales, we believe that it helps to clarify the essential underlying mechanism,
the intuition and important figures of merit behind equilibration.
We then connect our arguments to common assumptions and numerical results in the field of equilibration and thermalization of closed quantum systems, such as the eigenstate thermalization hypothesis as well as rigorous results on interacting quantum many-body systems.
Finally, we complement our discussion with numerical results --- both in the case of examples and counter-examples of equilibrating systems.
\end{abstract}

\maketitle
\section{Introduction}

How do closed quantum many-body systems precisely equilibrate, once pushed out of equilibrium? This question has intrigued already the pioneers of quantum mechanics \cite{vonNeumann29}. Early on, it became clear that generically, expectation values of local observables eventually become stationary, so that the systems would appear relaxed for most times, even though undergoing perfectly unitary dynamics globally. Effectively, a dephasing mechanism is seen at work that ensures that for most times, the state is locally indistinguishable from the time average. It was rather recently that this intuition was made rigorous by a solid quantitative understanding of equilibration living up to the expectations of mathematical physics \cite{christian_review,Linden_etal09,1402.1093,Equilibration2,ReimannNC,Wintereq,CramerEisert,ReturnToEquilibrium}, backed up by a large body of numerical work on interacting quantum many-body systems \cite{Sengupta_Silva_Vengalattore_2011,RigolRandomMatrix} and on integrable models \cite{CauxEssler,CauxIntegrable}. In particular, if the so-called effective dimension of the initial state is large, then local equilibration follows in a stringent sense \cite{Linden_etal09,1402.1093,Equilibration2,ReimannNC,Wintereq,CramerEisert,ReturnToEquilibrium}.
In  fact, the study of quantum many-body systems going back to equilibrium has seen a renaissance in recent years, dubbed quantum systems undergoing ``quenches'' \cite{CalabreseCardy06,Rigol_etal08,CramerEisert,christian_review,1408.5148,Sengupta_Silva_Vengalattore_2011,Yukalov2011,RigolRandomMatrix,1408.5148}, 
not the least because such systems can now be probed in modern laboratories under well controlled conditions \cite{BlochSimulation,1408.5148,nature_bloch_eisert,RoschTransport,Gring_etal12}.

In all this deepened understanding, an important question has been left notoriously open, however. This is the question of how quickly such an equilibration would happen in time. This question is often referred to the ``time scale problem'': It is a frequent topic of sessions highlighting important open questions in the field (even though the term ``scale'' may be a misnomer, as one does not necessarily expect local relaxation to follow
an exponential law in time). \emph{General} bounds on equilibration time scales have been proven \cite{Linden_etal09,1402.1093,Equilibration2,Wintereq}, but they scale very much unfavorably with the system size, while one would expect the dependence in time to be largely independent from it. For non-interacting models, the question of equilibration
is much better understood in contrast \cite{CramerEisert,Marek,CalabreseEsslerFagotti11,Farrelly2016,CramerEisertScholl08}. 
But then, such systems show distinct features generic non-integrable models would not exhibit.

The question of the equilibration in time for local Hamiltonian models is the one discussed in the present work. Yet before we state what this work can deliver, we take the opportunity to clarify what this work is not. We do not present a rigorous proof of equilibration times. We do not even present bounds to such times for specific Hamiltonians. Instead, in this work, we aim at providing a clear intuition for equilibration times based on physical principles. We make this intuition plausible by formulating basic assumptions about interacting quantum
many-body systems and link those to observable equilibration processes by making use of a simple argument from harmonic analysis. This serves to highlight important quantities that determine whether an observable equilibrates relative to a state and some Hamiltonian or not.
We then put those principles advocated and assumptions made in contact with rigorous results on interacting quantum systems and to commonly conjectured properties. To further add evidence to the plausibility of the assumptions, we present a body of numerical observations which illustrate the different possible behaviors.

We regard this work hence largely as a discussion document on an important research question, with the hope of providing some intuition and inviting interested readers to work on this interesting problem. We are largely unapologetic about this format.
In fact,
we think that it is very reasonable for the difficult and long-standing problem at hand to choose such a format in order to communicate and further develop the necessary intuition.

\section{The notion of equilibration}
We will now start off by explaining what we mean by equilibration in a many-body system and what kind of phenomena we are interested in.
Generally speaking, the phenomenon of equilibration means that if a (quantum) system described by a Hamiltonian $H$ is prepared in a state $\Psi$ and we measure an observable $A$ at different times, we find that the expectation value becomes largely time-independent
\begin{align}
\langle A(t)\rangle_\Psi \rightarrow \mathbb{E}_t(\langle A(t) \rangle_\Psi) = \mathrm{const.}
\end{align}
Here, the quantity $\mathbb{E}_t(\langle A(t) \rangle_\Psi)$ denotes the infinite time average of $\langle A(t) \rangle_\Psi$
\begin{align}
   \mathbb{E}_t(\langle A(t) \rangle_\Psi) = \lim_{T\rightarrow \infty} \frac{1}{T}\int_0^T \langle A(t) \rangle_\Psi\, \rmd t.
\end{align}
Whether, and if so how quickly, such equilibration happens depends crucially on the Hamiltonian $H$, the initial state $\Psi$ and the observable $A$. If we vary these parameters, we can in principle describe a vast set of complex phenomena, ranging from the equilibration of local order parameters, over the equilibration of macroscopic bodies at different temperatures which are put into contact, to the evaporation of a black hole by Hawking radiation.
To have any hope to make progress, we therefore have to formulate precisely in which kind of phenomena we are interested in and in which kind of phenomena we are not interested in.

In this work, we are interested only in the microscopic, local equilibration of
local observables in a sufficiently homogeneous quantum many-body system with
local interactions. That is, we want to provide plausible arguments for the
empirical fact \cite{nature_bloch_eisert} that in large interacting many-body
systems, local observables rapidly equilibrate into a \emph{local} equilibrium.
After this local equilibration has happened throughout this system, different
parts of the system might still be out of equilibrium on macroscopic length and
time scales. We will not be interested in this remaining equilibration, which
can be expected to be described by a semi-classical, hydrodynamic approach
\cite{Spohn1991}.

Thus, we exclude in our discussion essentially all effects originating from the system to be out of equilibrium on a
macroscopic scale.
This includes all conduction or transport effects on
macroscopic scales, such as, for example, heat conduction or electric currents.
This might seem like an overly strong restriction, but we emphasize that even this limited range of phenomena that we would like to describe is at the moment not thoroughly understood from a theoretical point of view \cite{1402.1093}. In particular, it has not been proven on general grounds that such local equilibration happens within a time that does not diverge with the system size of the whole system (see Refs.\ \cite{Wintereq,ReimannNC} for some rigorous progress in this direction, however).
Contrarily, we will argue that the essential mechanism of equilibration in a time independent of the system size is in fact quite simple. Indeed it is not significantly more difficult to understand than the spreading of the wave packet of a particle on a line. Nevertheless, it seems very challenging
 to obtain rigorous proofs for bounds on equilibration time-scales, at least if one aims at deriving those bounds directly starting from the microscopic model.

We also emphasize that we are only interested in the question of how rapidly local observables in a somewhat homogeneous system equilibrate to a steady-state value, but not what this value is. That means we are not concerned with the problem of whether this steady-state value can be described well by a statistical ensemble like the Gibbs-ensemble. As is common, we will refer to this latter phenomenon as \emph{thermalization} instead of equilibration (again, for reviews on this question, see Refs.\
\cite{Sengupta_Silva_Vengalattore_2011,1402.1093,christian_review, RigolRandomMatrix}).

\subsection{Basic notation and assumptions}
Before discussing the very essential mechanism of equilibration, we will now start to formalize our set-up by establishing some basic notation. Later in the manuscript, we will iteratively refine our formal set-up and assumptions. In the following, we will consider a system on a regular lattice $\Lambda\subseteq \ZZ^D$,  described by a local Hamiltonian
\begin{equation}
	H_\Lambda = \sum_{X\subset \Lambda} h_X,
\end{equation}
where the operator $h_X$ only acts on degrees of freedom within the lattice region $X$. Here, we assume that the degrees of freedom associated to each lattice site $x\in\Lambda$ are described by a finite-dimensional Hilbert-space, with local dimension $d$. Furthermore, we assume that the terms $h_X$ are non-zero only for regions $X$ of a finite diameter smaller than or equal to $k$ and are uniformly bounded in the sense that $\norm{h_X}\leq J$ for all $X\subset \Lambda$. We label energy eigenvectors as $\ket{E_i}$, with $i=1,\ldots,d^N$ and assume that the ground state energy is zero, so that $E_i\in[0,\infty)$ for all $i$.

Local observables are denoted as $A$. Without loss of generality, we pick $\|A\|=1$. We denote the number of terms $h_X$ as $n$ and the total number of sites by $N=|\Lambda|$. Usually, $n=N$.
Throughout this article, we will assume that the Hamiltonian $H_\Lambda$ is translational invariant, at least on the scale of the observables that we consider. Similarly, we will only allow for initial states which are translational invariant. Let us emphasize, however, that the reason for this is not that we do not believe that systems which are not strictly translational invariant would not equilibrate.
In fact, the basic picture of local, rapid equilibration also applies to these cases based on the following argument.
If we know that a local observable in a translational invariant system equilibrates in time $t_{\mathrm{eq}}$, the \emph{Lieb-Robinson bounds}
(see Sec.~\ref{sec:lieb-robinson}) imply that the local system essentially has only seen a system of size on the scale $v_{\mathrm{LR}}t_{\mathrm{eq}}$, where $v_{\mathrm{LR}}$ is the maximum group velocity implied by the Lieb-Robinson bounds. We can hence expect that results on equilibration in translational invariant systems are also relevant for the local, rapid equilibration of systems which are inhomogeneous on scales larger than $v_{\mathrm{LR}}t_{\mathrm{eq}}$.
An assumed separation of length-scales of non-equilibrium effects would therefore directly translate to a separation of equilibration time-scales while the initial (fast) equilibration process is the one we address here.
For example, many-body localized systems equilibrate in spite of breaking translational invariance
 \cite{Schreiber842,PhysRevB.90.174302}. Moreover, in our
numerical calculations presented in the main text and appendix we also consider non-translational invariant systems which still equilibrate and follow the laid out intuition.
However, translational invariance can be used to disentangle the problem of local equilibration with the problem of equilibration on macroscopic scales: If local order parameters equilibrate in a translational invariant system, then there is no remaining macroscopic non-equilibrium.

In the remaining document we will often make statements about different system sizes. Assuming a translational invariant state and Hamiltonian the system size can be changed unambiguously and without much effort.

\section{The essential mechanism: Equilibration of complex numbers}
\label{sec:The essential mechanism: Equilibration of complex numbers}
\begin{figure*}[t]
\centering
\includegraphics[width=.8\textwidth]{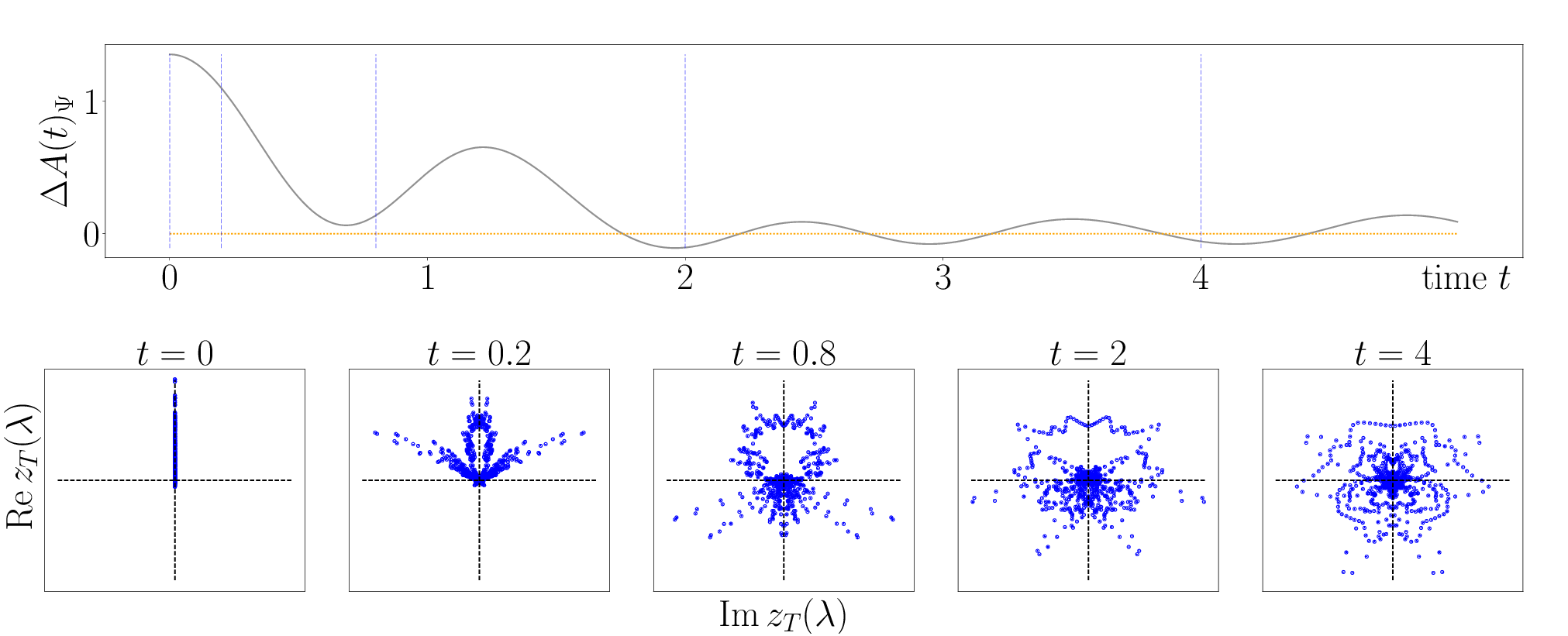}
\caption{Time evolution in an equilibrating system.
We show the exact time
evolution of the deviation of the instantaneous expectation value of a local
observable with respect to the steady-state value.
The model is the
transverse field Ising defined in Eq.~\eqref{eq:AppTransIsing} in the appendix on $L=15$ sites with parameters $J=4,h_x=1,h_z=-2.1$.
The initial state is a random product state with a spin-up state in the middle
and the observable is a $\sigma^z$ operator on that very spin as used
in Ref.\ \cite{PollmannTDVP}. A finite size analysis can be found in the
appendix.
Moreover, in the lower
panel, we plot the evolution of the individual terms contributing to the Fourier transform of the distribution $z_T$ in the complex
plane.
For numerical reasons $z_T(\lambda)$ is thereby evaluated at $5000$ points $\lambda_i$ which linearly interpolate the minimal and maximal gap $\Delta_{\mathrm{min}}$ and $\Delta_{\mathrm{max}}$ occurring for the respective system size with $T\approx 33$.
The discretization of $z_T$ is hereby ensured to approximate the regularized function well and we verify for instance
that $\sum_i z_T(\lambda_i)$ agrees up a relative error of $10^{-8}$ with the integral over $\int z_T(\lambda) \mathrm{d}\lambda = \sum_{\Delta\neq 0} z_\Delta$  and therefore no weight of the distribution is lost.
 We then plot the values of $z_T(\lambda_i)e^{\ii\lambda_i t}$ at the times marked in the evolution in the upper panel.
For the computation of $z_T(\lambda)$ we discarded all values $z_\Delta$ with $|\Delta|<10^{-13}$ in order to account for the subtraction of the infinite time averaged expectation value.
While initially strongly localized and anisotropic,
the time evolved $z_T(\lambda_i) e^{\ii \lambda_i t}$ quickly relaxes
into an isotropic distribution that is up to minor fluctuations
constant in time.
The plot of the distribution $z_T$ over $\lambda$ shown in Fig.~\ref{fig:zdPollmann} accordingly shows a large number of different gaps (and therefore angular velocities) that carry about equal weights of the distribution $z_T$.
This is directly reflected in the time evolution of
the deviation from the steady-state expectation value which decreases in time as the distribution spreads.}
\label{fig:EQtime}
\end{figure*}

\begin{figure*}[t]
\centering
\includegraphics[width=.8\textwidth]{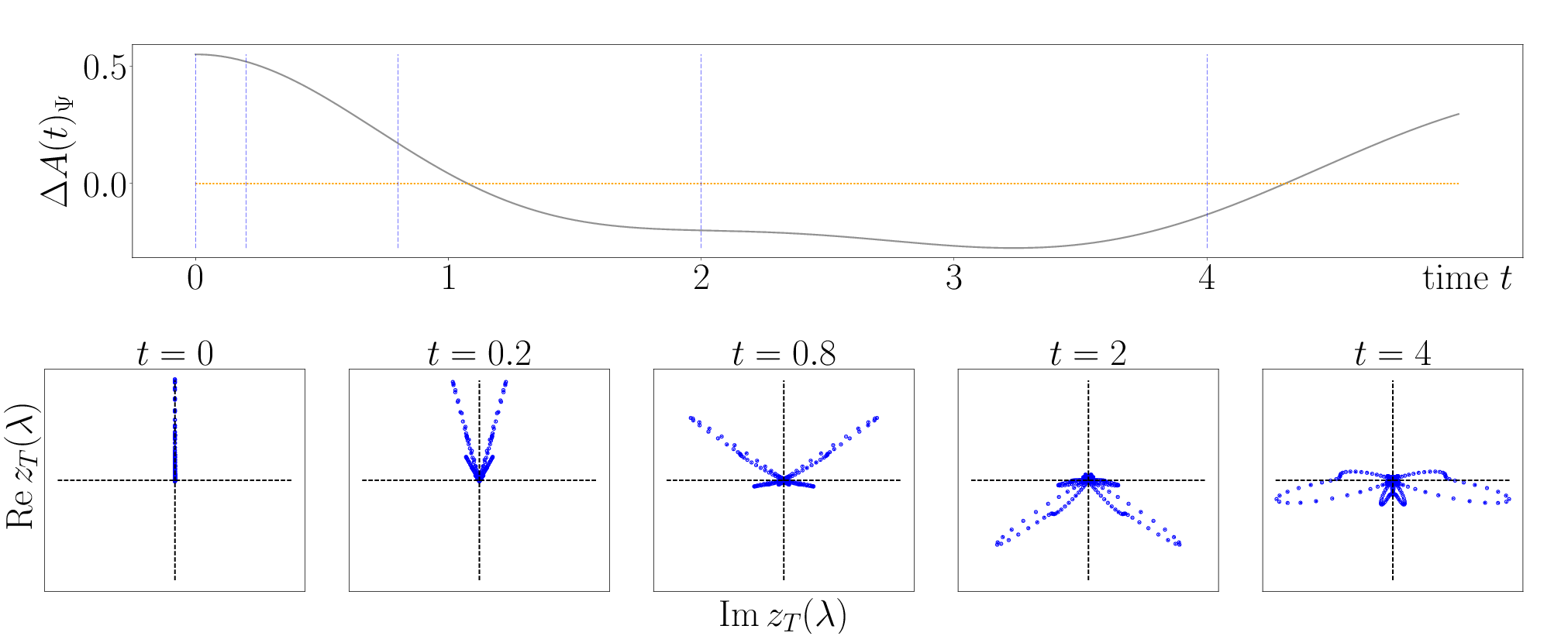}
\caption{Time evolution in a system failing to equilibrate. The
system size is $L=15$.
We show the exact time
evolution of the deviation of the instantaneous expectation value of a local
observable with respect to the steady-state value.
The model is the
transverse field Ising defined in Eq.~\eqref{eq:AppTransIsing} in the appendix on $L=15$ sites with parameters $J=1,h_x=0.5,h_z=-1.05$.
The initial state is a product state composed of only spin-up states
and the observable is a $\sigma^z$ operator in the middle of the chain as used
in Ref.\ \cite{Banuls}.
Moreover, in the lower
panel, we plot as in Fig.~\ref{fig:EQtime} the evolution of the contributions to the Fourier transform of the smoothed distribution $z_T$ in the complex
plane at the times marked in the evolution. We apply the same scheme as described in the caption of Fig.~\ref{fig:EQtime} again with $5000$ interpolation points, $T\approx 33$ and treating gaps $|\Delta|<10^{-13}$ as zero.
Initially, the smoothed $z_T$ is strongly localized and anisotropic. When
evolved in time, we find two distinct and large contributions that revolve
around the zero without canceling out one another. These contributions stay approximately in phase and do not disperse as their parts revolve with roughly the same angular velocity.
Note that this agrees well with the result shown in Fig.~\ref{fig:zdBanuls} which displays the distribution $z_T$ in dependence of $\lambda$ and shows two distinct peaks concentrating most of the weight of the distribution.
As a result the deviation from the steady-state expectation value shows strong and only weakly decaying oscillations.}
\label{fig:NEQtime}
\end{figure*}
We will now explain the very essentials of the mechanism of local equilibration in a complex quantum system. The point of view that we will use is also advocated in the 
concurrent and complementary Ref.\ \cite{ArnauEquilibration} 
which also deals with the problem of rapid local equilibration. 
To explain this mechanism, which is essentially that of simple \emph{dephasing}, we fix the initial
pure state $\Psi$ and consider the quantity
\begin{align}
 \DelA{t} := \langle A(t)\rangle_\Psi - \mathbb{E}_t(\langle A(t) \rangle_\Psi),
\end{align}
which measures the deviation of the instantaneous expectation value of the observable $A$ from the steady-state value. It is instructive to write this quantity in the energy-eigenbasis as
\begin{align}
  \DelA{t} &= \sum_{E_i\neq E_j} \bra{E_i}A\proj{E_j}\rho\ket{E_i}\e^{\ii (E_j-E_i)t}\nonumber\\ &= \sum_{\Delta\neq 0}z_\Delta \e^{\ii \Delta t}, \label{eq:fundamentalequation}
\end{align}
where we have introduced a sum over the gaps of energy eigenvalues $\Delta$ and the complex numbers
\begin{align}
  \label{eq:zdelta}
  z_\Delta := \sum_{\substack{E_i,E_j\\ E_i-E_j=\Delta}} \bra{E_i}A\proj{E_j}\rho\ket{E_i}.
\end{align}
Due to the fact that $A$ is hermitian and the time-evolution unitary, the
$z_\Delta$ fulfill the relation $z_{-\Delta} = \overline{z_\Delta}$, where the
bar denotes complex conjugation.

We will now use this expression to give an intuitive understanding of
equilibration. In a large many-body system, the number of gaps $\Delta$ grows
exponentially with the system size. For sufficiently generic local observables and
initial states we then expect that the number of points $z_\Delta$ that
contribute to $\DelA{t}$ is very large in a large system. From
Eq.~\eqref{eq:fundamentalequation} we then see that the time-dependent deviation
$\DelA{t}$ can be understood as the sum of a cloud of a large number of
points in the complex plane, each of which rotating on a circle of radius
$|z_\Delta|$ and with an angular velocity given by $\Delta$ (e.g.~see lower panel of
Fig.~\ref{fig:EQtime}). No two points have exactly the same angular velocity and there will therefore necessarily be a dispersion in the angular velocities.

If $|\DelA{0}|\gg 0$, the initial distribution of points is anisotropic and the dispersion in the angular velocities will have the effect that this initial anisotropy evens out, i.e., the points will start to distribute more isotropically in the complex plane. This leads to a small value of $\DelA{t}$. Once the points are spread out approximately isotropically, they will remain approximately isotropic for a long time: Intuitively speaking, there are vastly more configurations for the cloud of points so that they remain isotropic than configurations which lead to a sudden synchronization again. We hence expect that the points remain isotropically distributed for a long time. Nevertheless, in any finite system there will be a recurrence time \cite{Wallace2013}, which, however, increases quickly with the number of points.
In a local system of finite size, there is an additional recurrence-like time
due to ballistic transport of information and backscattering at the boundaries
of the system which increases as the system size, see Sec.~\ref{sec:lieb-robinson}.
In contrast, the time $t_{\rm eq}$ it takes until the points have distributed roughly isotropically, i.e., $\DelA{t_{\rm eq}} \leq \epsilon$, intuitively depends essentially on the shape of the distribution $z$ as a function of $\Delta$ and should become independent of the system size for large enough systems.
After this time, there will be remaining oscillations with a small amplitude $\epsilon$ roughly until the recurrence time. The size $\epsilon$ of the oscillations decreases with the number of points, while the recurrence time increases with this number. These arguments hold provided that the distribution of points is essentially fixed and sufficiently regular as we increase the number of points.

As a toy model for this, let us consider what happens if we choose a large number $N$ of gaps $\Delta_i$ uniformly at random from an interval $[-\Delta_{\rm max},\Delta_{\rm max}]$.
Let us furthermore simplify the situation by assuming that the $z_{\Delta_i}$
are real and distributed according to a Gaussian density $\mu(\Delta)$
(independent of $N$) with variance $1/\tau\ll \Delta_{\rm max}$.
We normalize their sum so that
\begin{align}
  \sum_i z_{\Delta_i} = \DelA{0} \gg 0
\end{align}
is fixed
(Note that the condition that $\DelA{.}$ is always real can be violated in this toy example,
however, only on a scale that decreases with increasing $N$. We therefore ignore this for now.).
We then see that as we increase $N$, the time-dependent deviation $\DelA{t}$ becomes a better and better approximation of the Fourier transform of a Gaussian
\begin{align}
  \DelA{t} = \sum_{i} z_{\Delta_i}\e^{\ii \Delta_i t} \approx \int \mu(\Delta)\e^{\ii \Delta t}.
\end{align}
For any given error $\epsilon>0$, there will be a maximum time $T_N(\epsilon)$ for which this approximation holds true. For times $t<T_N(\epsilon)$ we then find
\begin{align}
 \DelA{t} \approx \DelA{0}\e^{-(t/\tau)^2},\quad t<T_N(\epsilon).
\end{align}
Importantly, the time $T_N(\epsilon)$ increases with $N$ for a fixed error $\epsilon$. We can hence identify the equilibration time-scale
as $\tau$, and find $t_{\rm eq}= C\tau$ for an equilibration up to some given precision defined by $C>0$.

In a more general situation, the distribution of $z_\Delta$ may have features at smaller scales than $1/\tau$. These features will equilibrate on correspondingly longer time-scales. But whenever the $z_\Delta$ follow a distribution that is essentially independent of $N$ for large $N$, the observable equilibrates in a time that is roughly independent of the system size, while the precision with which it does increases with the system size and the duration for which it remains equilibrated also increases with the system size.
The actual problem of explaining local equilibration is then to give arguments that render it plausible that the $z_\Delta$ are distributed essentially independent of the system size, while their number increases with the system size.

Unsurprisingly, to do this we will need to make additional assumptions about the system, the observable and the initial state and refine our notion of equilibration.
We will discuss both examples and counter-examples in well-known models (both integrable and non-integrable) in the following sections.

\begin{figure}
  \includegraphics[width=0.5 \textwidth]{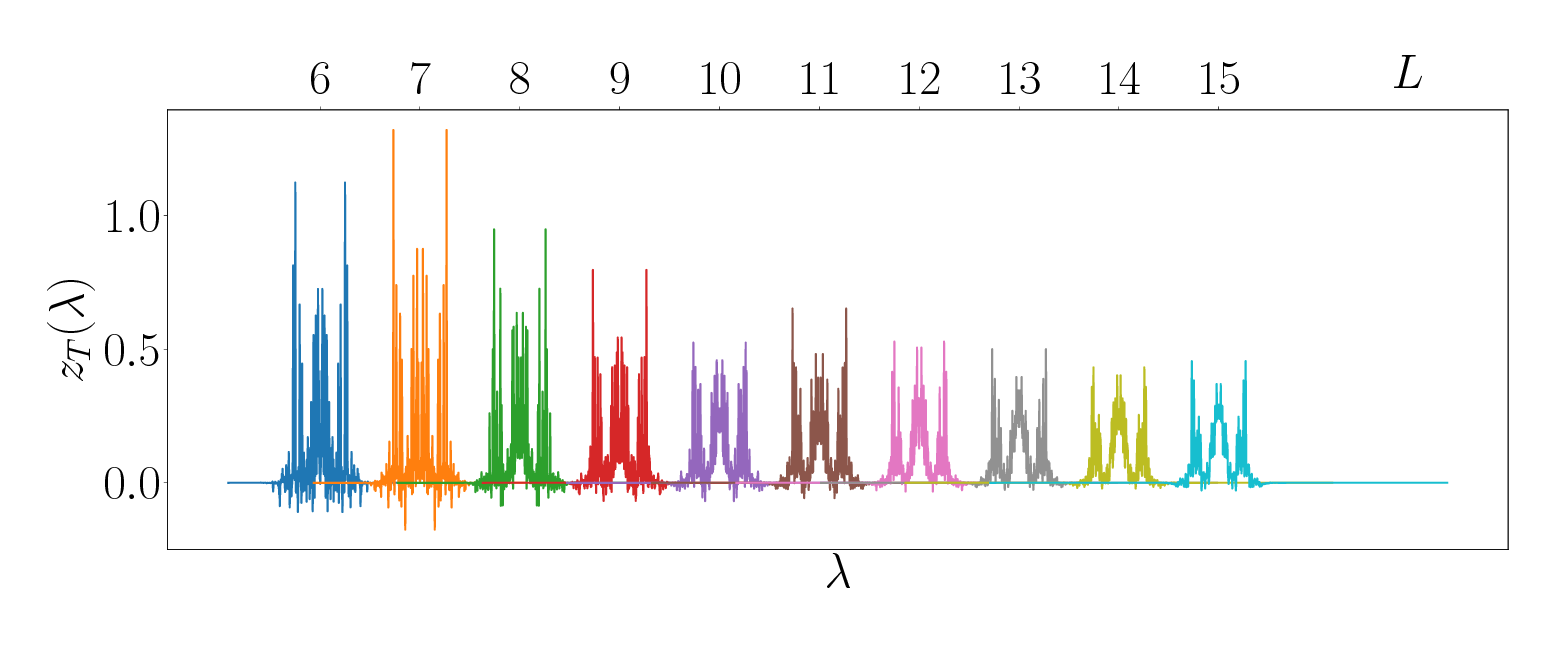}
  \caption{System size scaling of the smoothed $z_T$
  distribution in the equilibrating model discussed in Fig.~\ref{fig:EQtime}. At all system sizes the distribution $z_T$ is again evaluated at $5000$ points interpolating the extremal gaps linearly for $T\approx33$.
  With growing system size, the resulting distribution $z_T$ spreads its weights more and more evenly leading to the equilibrating behavior of the considered system as shown in Fig.~\ref{fig:EQtime}.}
  \label{fig:zdPollmann}
\end{figure}

\begin{figure}
  \includegraphics[width=0.5 \textwidth]{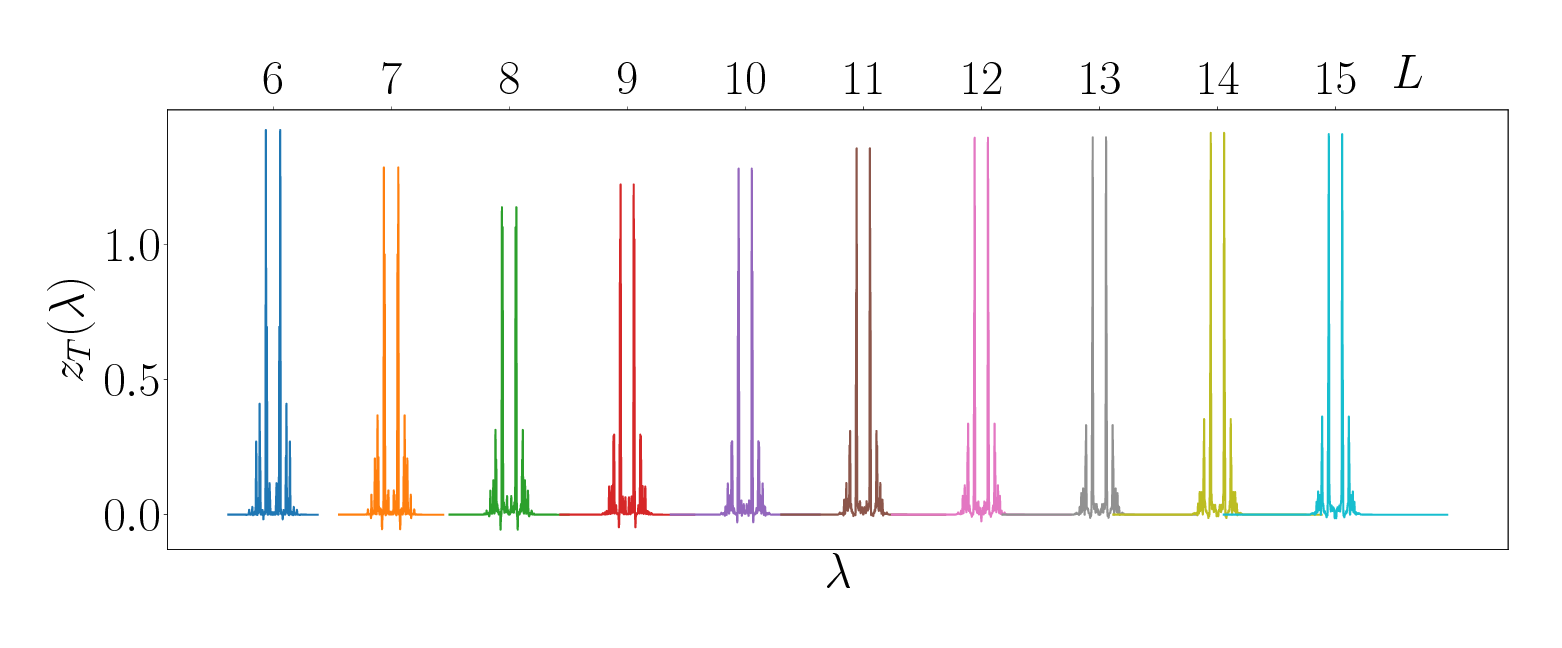}
  \caption{System size scaling of the smoothed $z_T$
  distribution in the non-equilibrating model discussed in Fig.~\ref{fig:NEQtime}.
  At all systems sizes the $z_T$ is evaluated at $5000$ points interpolating the extremal gaps $\Delta_\mathrm{min}$ and $\Delta_\mathrm{max}$ linearly for $T\approx33$. The resulting distribution $z_T$ concentrates most of the weight in two localized peaks which yield the non-equilibrating behavior shown in Fig.~\ref{fig:NEQtime}.}
  \label{fig:zdBanuls}
\end{figure}

%
%

\section{Revisiting basic notions and assumptions}
\label{sec:regularization}
In the previous section, we gave a rough intuitive argument for local
equilibration. We will now start to go into more detail and refine our notion
of equilibration. In the present section we will keep the discussion at a
general level and then connect this general discussion with more concrete
physical properties in the following sections.

%

We have already seen that in any finite system, there are different time-scales which need to be considered and that exact equilibration for infinite times will not occur due to the finiteness of the system: Not only will the system only equilibrate up to some finite precision,
 it will also only stay equilibrated for some very large but finite amount of time.
We should hence only ask whether equilibration occurs before some chosen cut-off time $T$ and up to some chosen precision $\epsilon$, which in general depends on the system size.
In the following, we will therefore only consider times $t<T$. Once we have introduced the cut-off time $T$ and a precision of equilibration
$\epsilon>0$, we can take the thermodynamic limit to obtain the time evolution $\DelA{t}$ in the infinite system.

Let us formalize this procedure to some extent. To do that, we introduce regularized quantities. We have seen before that the distribution of the $z_\Delta$ is the crucial quantity that governs the equilibration behavior. For any finite system, this is a discrete distribution of points. We will now regularize this into a smooth distribution while at the same time introducing the cut-off $T$. This is done by simply convoluting the distribution with a Gaussian of variance $1/T$. Let us therefore define the function
\begin{align}
 z_{T}(\lambda) = \sum_{\Delta\neq 0} z_\Delta \mc G_{1/T}(\lambda - \Delta),\label{eq:DefzTlam}
\end{align}
where $\mc G_\sigma$ is a normalized Gaussian of variance $\sigma$ and zero mean. Using the convolution theorem of Fourier analysis, we can then bound the instantaneous deviation $\DelA{t}$ as
\begin{align}
  |\DelA{t}| &\leq \left|\int z_T(\lambda)\e^{\ii \lambda t} \rmd \lambda\right| + \left|1-\e^{-(t/T)^2}\right| 2\norm{A} \\
                &\approx |\Delta A_T(t)_\Psi| + 2(t/T)^2 \norm{A},\quad t\ll T,
\end{align}
with
\begin{align}
\Delta A_T(t)_\Psi := \int z_T(\lambda)\e^{\ii \lambda t} \rmd \lambda.
\end{align}
Note that the regularized quantity $\Delta A_T(t)_\Psi$ always decays to zero on the large time-scale $T$.
It is thus essential to restrict the range of times that we are interested in to times much smaller than $T$.
We can now formalize the statement that the distribution of $z_\Delta$ becomes essentially independent of $T$ by saying that there exists a bounded function $\lambda\mapsto z(\lambda)$ such that
\begin{align}
  \label{eq:smoothlimit}
  \lim_{T\rightarrow \infty}
  \lim_{N\rightarrow \infty}
   \int |z_T(\lambda) - z(\lambda)| \rmd \lambda = 0.
\end{align}
In such a case, we will give strong arguments for equilibration of local observables under the evolution of a local Hamiltonian faster than
any power for any fixed precision in Sec.~\ref{sec:fourier_local_obersvable}.
For an example of such behavior, see Fig.~\ref{fig:EQtime} and \ref{fig:zdPollmann}.
In Fig.~\ref{fig:EQtime} we display the evolution of $\DelA{t}$ for an equilibrating system and
illustrate the connection to the intuition drawn in Sec.~\ref{sec:The essential mechanism: Equilibration of complex numbers}
for dynamic emergence of an isotropic distribution of $z_\Delta$. Fig.~\ref{fig:zdPollmann} shows the
behavior of $z_T$ for different system sizes and hints towards the development of an well behaved $z$ distribution in this example.

If, however, $\lambda\mapsto z(\lambda)$ is not point-wise bounded, we will now argue that the system cannot equilibrate at all to arbitrary precision: There will be remaining oscillations with finite amplitude for all times.
To see this, first note that for any system size and any $T$ we have
\begin{align}
  \DelA{0} = \sum_{\Delta \neq 0}z_\Delta = \int z_T(\lambda) \rmd \lambda.
\end{align}
Thus, by assumption $\lambda\mapsto z(\lambda)$ is an unbounded function with a finite integral. Such a function will have a finite contribution from the respective singularities which concentrate a finite weight onto arbitrarily small regions an thus lead to a non-dispersing evolution.
In the easiest case assume this contribution to originate from a discrete number of $\delta$-distributions.
Since $z_T(\lambda)=\overline{z_T(-\lambda)}$, they would have to come in pairs $\{+\Delta_i,-\Delta_i\}$ and hence show up in the time-evolution of the infinite system as
\begin{align}
  \DelA{t} \overset{t\rightarrow \infty}{\longrightarrow} \sum_i r_{i} \cos(\Delta_i t),
\end{align}
with $r_i$ being some real numbers. An example of a system which seems to show this
behavior is given in Figs.~\ref{fig:NEQtime} and \ref{fig:zdBanuls}, where two pronounced peaks in the $z_\Delta$ distribution lead to the non-equilibrating behavior displayed. Importantly, the example features a translational invariant, local Hamiltonian far from an integrable point.
Furthermore, the initial state is a translational invariant, product state and the observable is a on-site Pauli-observable. We note however, that there is some debate about whether these oscillations do indeed persist for all times in the infinite system or can be understood in terms of effective quasi-particles with a finite, but very long life-time  \cite{Slowest,Lin2017}.

As a final remark in this section, let us briefly discuss the role of
finite-size effects for the quantity $\DelA{t}$ which are necessary to keep in
mind in numerical investigations of equilibration on finite systems. If an
observable equilibrates in a time $t_{\mathrm{eq}}$ (w.r.t. some suitably chosen state) to the equilibrium value $A_{\mathrm{eq}}$ in the thermodynamic limit, it follows from the maximum group velocity $v_{\mathrm{LR}}$ implied by the Lieb-Robinson bounds that the observable should also equilibrate on a finite system of size $~v_{\mathrm{LR}}t_{\mathrm{eq}}$ to $A_{\mathrm{eq}}$ for some range of times. However, the equilibrium value $A_{\mathrm{eq}}$ need \emph{not} coincide exactly with the time-average value $\mathbb{E}_t(\langle A(t) \rangle_\Psi)$ on the finite system. This happens due to the fact that the latter may depend on the system size while $A_{\mathrm{eq}}$ by definition does not depend on the system size. To see examples of this behavior, we refer to the appendix.

\subsection{Rapid equilibration: A simple argument from harmonic analysis}
\label{sec:fourier_local_obersvable}

We will now give a simple argument showing that rapid equilibration, in the sense that it occurs in the thermodynamic limit and hence independent of the system size, essentially follows from the fact that $\lambda\mapsto z(\lambda)$ exists as a bounded function.

To be able to show this, we will use the fact that local observables can only connect energy-eigenstates which differ in a small amount of energy. More precisely, if $H$ is a local Hamiltonian with energy eigenvectors $\ket{E_i}$, we have
\begin{align}
  \label{eq:simplelocalglobal}
  |\bra{E_i} A \ket{E_j}| \leq \norm{A} \e^{-\alpha (|E_i-E_j| - 2R)},
\end{align}
where $R$ and $\alpha$ are constants independent of the system size and $R$ is
proportional to the size of the supporting region of $A$. This result has been
shown in ref.~\cite{LocalGlobal}.

In the case of a generic, strongly-interacting Hamiltonian we expect that the gaps $\Delta$ are essentially non-degenerate, meaning that for each gap $\Delta$ there are only few pairs of energies $E_i,E_j$ such that $E_i-E_j=\Delta$. In this case,
we see from Eqs.~\eqref{eq:simplelocalglobal} and \eqref{eq:zdelta} that for large $\Delta$, all $z_\Delta$ fall off exponentially with $\Delta$. This happens independently of the system size. Therefore, if $\lambda\mapsto z(\lambda)$
exists as a bounded function, it must also fall off exponentially with $|\lambda|$.
We thus find that $\lambda\mapsto z(\lambda)$ has the following properties: \begin{itemize}
\item[i)] It is bounded (by assumption), \item[ii)] it has a finite integral and \item[iii)] it decays exponentially with $|\lambda|$.
\end{itemize}
These three properties together imply that
\begin{align}
  \norm{z}_1 = \int |z(\lambda)|\, \rmd \lambda < \infty.
\end{align}
We will now essentially follow the proof of the Riemann-Lebesgue Lemma \cite{RiemannLebesgue}. Every absolutely integrable function can be approximated, to arbitrary small error $\delta>0$, by a smooth function with compact support $g_\delta$,
\begin{align}
\int |z(\lambda)-g_\delta(\lambda)|\, \rmd \lambda  < \delta.
\end{align}
Consequently, the Fourier-transform of $z$, which is simply $\DelA{t}$, can be approximated by the Fourier-transform of $g_\delta$,
\begin{align}
  |\DelA{t}| &\leq \left|\int g_\delta(\lambda)\e^{\ii \lambda t}\,\rmd \lambda\right|+\delta \nonumber\\
  &= |\hat{g}_\delta(t)| + \delta.
\end{align}
Since the Fourier transform of a compactly supported smooth function falls off faster than any power, we find that for every choice of precision $\delta$, $
t\mapsto \DelA{t}$ equilibrates to this precision faster than any power. For every $k\in\NN$ and $\delta >0$, there exist constants $C_k(\delta,\tau)$ such that
\begin{align}
  |\DelA{t}| \leq \min\left \{2\norm{A}, \frac{C_k(\delta,\tau)}{t^k}\right\} + \delta.
\end{align}
We emphasize, however, that the prefactors $C_k(\delta,\tau)$ can be very big and in fact diverge as $\delta\rightarrow 0$. This will, for example, be the case if the equilibration occurs with a power law behavior for large times, which is known to occur in certain integrable models
\cite{CramerEisert,Marek,CalabreseEsslerFagotti11}.
Nevertheless, the prefactors do not depend on the system size.
We therefore do not predict exponential laws in time but show that under the above assumptions system size independent equilibration times can be constructed.

%

\section{Connecting the assumptions to general results about quantum many-body systems}

The previous discussion rests on the assumption that the distribution $z_\Delta$ generically approaches a smooth distribution in the thermodynamic limit, in the sense of Eq.~\eqref{eq:smoothlimit}.
In this section, we connect the above assumptions with commonly stated and discussed properties of local quantum many-body systems.
As is clear from the expression for $z_\Delta$, the crucial properties that we will be interested in are
I) the distribution of energy gaps and II) how local observables and physically relevant initial states look in the energy-eigenbasis.
Finally, we will also discuss the role of a finite group velocity for the problem of equilibration.

\subsection{The spectrum}
As a preliminary for the following discussions,  the number of energy levels in a system of $N$ degrees of freedom
and local dimension $d$ is given by $d^N$, thus grows exponentially with the system size. In contrast, the energy-range of a local Hamiltonian grows only \emph{linearly} with the system size, $\norm{H}\propto N$. Thus,
it is reasonable that the typical energy difference between consecutive energy levels becomes exponentially small in the system size, leading to essentially a continuous spectrum for very large systems. As a consequence, also the distribution of energy gaps $\Delta$ follows a continuous distribution.

It can be seen that for systems with local interactions, the distribution of
energy levels follow roughly a Gaussian distributions with a standard deviation
that diverges as $\sqrt{N}$ with the system size. Precise results and a rough
intuition for why this happens can be found in the appendix (see Sec.~\ref{sec:rigorous:spectrum}). As a consequence of the distribution of energy levels, also the distribution of all energy gaps $\Delta=E_i-E_j$ has to follow a Gaussian distribution.

As a remark, we emphasize that it is important to distinguish the distribution of energy gaps $\Delta$ from the level-spacing distribution as
commonly studied in
\emph{random matrix theory}. In that case one is often interested in the distribution of \emph{neighboring} (or higher order) gaps and not on the distribution of \emph{all} gaps, as is necessary for the problem of equilibration. An important feature of the level-spacing distribution is that in sufficiently non-integrable models it shows \emph{level repulsion}. Nevertheless, the total distribution of gaps converges to a Gaussian with mean zero, i.e., a distribution with a maximum at the zero gap.

\subsection{The state}
The second important ingredient in the distribution of $z_\Delta$ is the initial state $\Psi$, and more concretely, the off-diagonal elements in its density matrix $\rho$. Our heuristic arguments require that this distribution is sufficiently smooth and dense as the system size increases. Due to the positivity of quantum-states one can easily bound their off-diagonal elements by their diagonal elements, according to
\begin{align}
  |\bra{E_i}\rho\ket{E_j}|^2 \leq \bra{E_i}\rho\ket{E_i}\bra{E_j}\rho\ket{E_j}.
\end{align}
At least if the state is sufficiently non-diagonal in the energy-eigenbasis, we can expect that there are many non-zero entries $\bra{E_i}\rho\ket{E_i}$, namely a number that diverges as the system size increases. Since they have to add up to unity, most of them will be quite small.

In the appendix (Sec.~\ref{sec:rigorous:state}) we will present general and rigorous results for the case of product states and more general states with a finite correlation length. These results show that the energy-distribution $\bra{E_i}\rho\ket{E_i}$ falls off (sub-)exponentially
for energies that differ macroscopically from the mean value, i.e., for energies that differ by more than a sublinear function in the system size from the mean. Indeed, for product states one can even proof a Gaussian decay.

For states exhibiting exponentially decaying correlations the divergence of the number of non-zero $\bra{E_i}\rho\ket{E_i}$ entries can in fact be argued rigorously.
Let us invoke the
 common definition of the \emph{effective dimension} of a given state $\rho$ as \cite{Linden_etal09,1110.5759}
 \begin{align}
   d_{\mathrm{eff}}(\Psi) = \frac{1}{\sum_{E_i} \bra{E_i}\rho\ket{E_i}^2}.
 \end{align}
The effective dimension measures the participation ratio of energy levels in
the initial state. It is rigorously known that whenever this quantity is very
large compared to the Hilbert-space dimension of the local system and there are
not too many degeneracies in the energy gaps, the local system equilibrates
to high precision after some unknown (and usually ``unphysically large'') time if the total system is very large but finite \cite{Linden_etal09}.
Recently, lower bounds for the effective dimension have been proven rigorously. If a state has  exponentially decaying correlations, then the effective dimension is lower bounded as \cite{ReturnToEquilibrium}
\begin{align}
  \label{eq:effective_dimension_lower_bound}
  \frac{1}{d_{\mathrm{eff}}(\Psi)}  \leq C \frac{\ln^{2d}(N)}{s^3\sqrt{N}},
\end{align}
where $C>0$ is a constant independent of the system size and
$s=\sigma/\sqrt{N}$. Here, $\sigma$ is the standard-deviation of the energy of
$\Psi$. The quantity $s$ is upper bounded independent of the system size, but
in many cases can also easily be lower-bounded. In particular, as was pointed
out in Ref.\ \cite{ReturnToEquilibrium}, for initial states that are product
states, one can compute $s$ explicitly, since only system-sizes of twice the
support of the interaction terms are necessary to be computed in a
translational invariant system. We thus see that Eq.~\eqref{eq:effective_dimension_lower_bound} shows that an ever-increasing number of energy levels participate the initial state.

It is important to note that these arguments only directly apply to the \emph{diagonal} entries of $\Psi$, whereas the $z_\Delta$ distribution depends only on the \emph{off-diagonal} entries. Note, however, that if we assume that the initial state is pure, as we do, a large number of small but non-zero diagonal entries also requires a correspondingly large number of non-zero off-diagonal entries to be compatible
with a unit rank of the density matrix.
These arguments thus support the hypothesis that for generic, strongly interacting models and initial states with short-range correlations, the distribution of $z_\Delta$ approaches a smooth distribution as the system size increases.

\subsection{The observable}
The last quantity that enters the distribution of $z_\Delta$ is the shape of a local observable in the energy-eigenbasis. As discussed before, we will see that local observables fall off exponentially on the off-diagonals. This can be shown rigorously.
Importantly, this fact implies that in the distribution of $z_\Delta$ only gaps of the size of the order of the support of the observable, and not the system size, are relevant. Since by the above mentioned results both the
 distribution of energy gaps and the probability distribution of energies have a standard deviation which grows sublinearly with the system size, we can hence assume that both distributions are effectively uniform on the relevant scale---namely the support of the observable.

 One of the most frequently invoked assumptions on interacting quantum many-body systems in the context of studying the \emph{thermalization} of closed quantum systems is the \emph{eigenstate thermalization hypothesis} (ETH) \cite{Deutsch91,Srednicki94,RigolRandomMatrix}. It basically assumes that when considering expectation values of local observables, individual eigenstates of $H$ are already expected to give predictions very similar to those of Gibbs states. This is generally seen as a highly plausible hypothesis, at least for eigenvectors of $H$ corresponding to eigenvalues
 in the bulk of the spectrum of strongly interacting many-body systems. There are several specific formulations of this hypothesis.

 If local observables of global energy-eigenstates should be given by the those of the thermal state with the same energy, it follows that the diagonal elements $\bra{E}A\ket{E}$ depend smoothly on the energy $E$. In addition, it is often argued that also the off-diagonal elements of $A$ obey a smooth distribution up to small fluctuations. One of the strongest
 forms of stating the eigenstate thermalization hypothesis is the following one
 as it is stated in the review Ref.\ \cite{RigolRandomMatrix}, even
 though several other formulations are known.

 \begin{assumption}[Eigenstate equilibration hypothesis \cite{RigolRandomMatrix}]
 \begin{align}
     \bra{E_i}A\ket{E_j} =  g(\bar E) \delta_{i,j} + e^{-S(\bar E)/2} f_A(\bar E,\omega) R_{i,j}
 \end{align}
 where $\bar E:= (E_i+E_j)/2$, $\omega=E_i-E_j$, and $E\mapsto S(E)$ is the
 micro-canonical entropy associated to energy $E$. $f$ and $F_A$ are both assumed to be smooth functions of their arguments, $f(\bar E)$ being the expectation value of the microcanonical ensemble at energy $\bar E$, and $R_{i,j}$ being a random (real or complex) variable with zero mean and unit variance.
 \end{assumption}

The ETH thus gives further plausibility to the assumption that the $z_\Delta$
approach a smooth distribution for large systems and initial states with energy
densities in the bulk of the spectrum. This is true even despite the
fluctuations $R_{i,j}$: In the limit of large systems they should average to
smooth distributions after we regularize with any Gaussian of finite width
(compare with Sec.~\ref{sec:regularization}).

\subsection{The role of Lieb-Robinson bounds}
\label{sec:lieb-robinson}
Before coming to the conclusion of this discussion paper, we would like to
discuss the role of a finite group velocity for the problem of equilibration.
It is known that that in every local quantum many-body system with
finite-dimensional local Hilbert-spaces, there is a finite velocity
$v_{\mathrm{LR}}$ with which information and excitations can spread through the
system---an effect akin to a light-cone in special relativity. Unlike in
special relativity however, the "light-cone" is not strict, but processes which
violate the "light-cone" condition are exponentially suppressed. This result is known as \emph{Lieb-Robinson bounds} (LR-bounds) and the corresponding "light-cone" is usually called the \emph{Lieb-Robinson cone} (LR-cone). The result was proven for the first time in Ref.\ \cite{LiebRobinson72} and one way to express it is as
\begin{align}
\norm{[A(t),B]} \leq c \norm{A}\norm{B}  \e^{-a(\mathrm{d}(A,B)-v_{\mathrm{LR}}|t|)},
\end{align}
where $A,B$ are local observables,
$a,c>0$ are constants and $\mathrm{d}(A,B)$ is the lattice-distance between the observables $A$ and $B$ (the above
stated bound can be tightened).
In many applications of LR-bounds in many-body physics, one can in fact practically neglect the tails of the LR-cone in exchange for an arbitrarily small error. As one would expect, the group velocity $v_{\mathrm{LR}}$ is essentially determined by the interaction strength. Also note that LR-bounds are independent of the initial state.

In the context of equilibration, the LR-bounds have the important effect that, since information can only travel with a finite velocity, the local observable $A$ only sees a small part of the full system. If it is necessary to sense a length-scale $l$ for the local system to equilibrate, the finite group velocity therefore puts a \emph{lower bound} on the equilibration as $t_{\mathrm{eq}}\geq l/v_{\mathrm{LR}}$.
%
%

Thus, LR-bounds tell us that in many cases we do not need to go to arbitrarily large system sizes when we do numerical checks of equilibration times. If we see rapid equilibration for small system sizes, we know that it will also be true for larger system sizes. On the other hand, if we can simulate a system up to linear system size $L$, then we should only consider the time-evolution up to a time of about $t=L/v_{\mathrm{LR}}$.
This is due to the fact that if we let the simulation run for a time longer than $L/v_{\mathrm{LR}}$ information can travel across the system and come back to the local system we are interested in. This leads to a similar effect as a recurrence time. Thus, only simulations up to times of the order of $L/v_{\mathrm{LR}}$ should be considered. If we do not see equilibration in such a simulation, we have to increase the system size. Lieb-Robinson bounds hence allow us to confirm but unsurprisingly not to falsify equilibration on small systems.

\subsection{A small comment on disordered systems}
In this article, we have always considered translational invariant systems.
As indicated before, we do believe that the simple arguments that we gave should be convincing also
for systems that deviate from this strict assumption. In our numerical calculations, we also show systems that fulfill translational
invariance and fail to equilibrate (see Fig.~\ref{fig:NEQtime}), as well as
systems that break translational invariance in the initial state but do
equilibrate (see Fig.~\ref{fig:EQtime}).
 Indeed, it is well known
that disordered, but strongly interacting systems---so-called \emph{many-body
localized} systems---also equilibrate \cite{Schreiber842,PhysRevB.90.174302}.
Such systems fulfill stronger bounds than the LR-bounds, in which the light-cone is deformed away from a linear cone into a logarithmic cone. Nevertheless, there is still a spread of information (but not of quasi-particles) throughout the system and dephasing  in the sense as we have explained it can happen \cite{1404.5216,1412.3073,1412.5605}.
In non-interacting disordered systems, like Anderson insulators, not even
information can propagate across length-scales larger than the localization
length \cite{HamzaSimsStolz12,Burrell2007}
Thus, local excitations essentially live in a small, finite system and no local equilibration is observed.

\subsection{Typicality  results}
We finally take the opportunity to relate the results laid out here to notions
of ``typical fast thermalization'' processes in closed many-body systems as
discussed in Ref.\ \cite{ReimannNC}. There, fast equilibration is also derived
based on physically meaningful assumptions, albeit quite different ones than
the ones discussed here. In the approach pursued there, the unitary $U$
transforming the eigenbasis of $A$, $\rho$ and that of $H$ is in the focus of attention. Akin to random matrix theory, it is argued that one should expect that the  overwhelming majority of randomly sampled $U$ should typically also apply to the particular, non-random actual system of interest and that the Haar measure serves as a meaningful probability measure here. Making that assumption based on physical plausibility, one can indeed derive rigorous bounds to equilibration times. In contrast, in the present work, the structure of the problem for local Hamiltonians is in the center of interest. We thus see our discussion as complementary to that in Ref.\ \cite{ReimannNC}, hoping to provide a more "mechanical" picture of how equilibration occurs.

\section{Discussion}
In this article, we have elaborated on a simple explanation how equilibration of local observables happens in strongly interacting many-body systems. Our aim was not quite to prove rigorous results, but to establish an intuition
why local equilibration is plausible and why it is also reasonable to expect that it happens independent of the system-size.
This explanation relies on many assumptions which in many systems seem most natural.
To give further substance to the plausibility of the assumptions,
we have connected to some extent our arguments with rigorous and conjectured general results about quantum many-body systems.
Moreover, we have furthermore demonstrated this behavior for specific numerical examples. We have put particular emphasis on the necessity to regularize numerical quantities. This is both necessary to meaningfully speak about equilibration in finite systems, as well as to make sense of numerical data.
Nevertheless, there are of course counter-examples of the discussed behavior,
such as the one studied in Ref.\ \cite{Banuls} and displayed in
Fig.~\ref{fig:NEQtime}. This is not surprising, due to the simplicity of our arguments and can indeed nicely be illustrated using the regularized quantities that we introduced.

When formulating this discussion paper, we have aimed at being as educational as possible with the hope that readers with a different background but interested in the problem can get a basic understanding of the mechanism of equilibration. Of course, much of the
problem of equilibration can still be considered largely unsolved: Key is to
identify general physical explanations for when local rapid equilibration happens, that is finding properties that allow to make predictions about strongly interacting models and their equilibration times without having to solve them, directly based on and derived from
the microscopic model at hand. We sincerely hope that this work
helps to motivate readers to investigate this intriguing problem in more detail.

{\bf{Acknowledgments}}: We acknowledge frequent discussions with colleagues about this
question over the years, specifically within the COST action on quantum thermodynamics.
Specifically, we would like to thank Arnau Riera and his co-authors for sharing the
manuscript of Ref.\ \cite{ArnauEquilibration} prior to publication.
We acknowledge funding from the DFG (CRC 183, EI 519/7-1, GA 2184/2-1), the BMBF, the
EU (AQuS), the Templeton Foundation, the ERC (TAQ) and the Studienstiftung des Deutschen Volkes.


\bibliographystyle{naturemag}

\appendix
\onecolumngrid
\section{Numerical implementation and investigations}
In this appendix, we present the numerical procedure to obtain the plots from
the main text. All calculations rely on full exact diagonalization for
system sizes $L \in \{6,\ldots,15\}$ and open boundary conditions.
The general procedure is the same for
all models considered. After diagonalizing the Hamiltonian $H$, we obtain
the eigenenergies $E_i$ and the diagonalizing unitary $U_d$. The latter is
then used to transform the state $\rho$ and the observable $A$ into the
energy eigenbasis of $H$. Note that in real space both might feature
some sparsity structure that one might exploit, however this structure vanishes after the transformation for generic Hamiltonians.
Given $\rho$ and $A$ in the energy eigenbasis as well as the energies $E_i$,
we are now set to calculate the exact and discrete $z_\Delta$
distribution according to its definition given in Eq.~\eqref{eq:zdelta}.
In order to subtract the steady-state value, we discard all
$z_\Delta$ for which $|\Delta| < 10^{-13}$.
The  discretization of the regularization $z_T$ is then calculated at $5000$
linearly spaced $\lambda_i$ which interpolate the extremal gaps $\Delta_{\mathrm{min}}$ and $\Delta_{\mathrm{max}}$ of the given problem using Eq.~\eqref{eq:DefzTlam}.
The discretization of $z_T$ is hereby ensured to approximate the regularized function well and we verify, for instance, that $\sum_i z_T(\lambda_i)$ agrees up a relative error of $10^{-8}$ with the integral over $\int z_T(\lambda) \mathrm{d}\lambda = \sum_{\Delta\neq 0} z_\Delta$  and therefore no weight of the distribution is lost.
Lastly, the time evolution we presented in the main text and here in the appendix are the time evolved
$\DelA{t}$ that are obtained by evolving the initial state
integrating the Schroedinger equation numerically and tracking the expectation
value $\langle A(t)\rangle_\Psi$. Here we simulated the
time interval $[0,5]$ in 1000 steps.

For the purpose of conveying the general idea of the importance of the
$z_\Delta$ distribution, we would like to present explicit calculations,
that are of course based on specific models. While in the main text we focus on two models, we actually investigated four models which
nicely contrast different equilibration behavior and have been considered
in previous works. We set out to connect their capability or incapability to equilibrate to
their $z_\Delta$ distribution underlining the intuition laid out in the main
text in Sec.~\ref{sec:The essential mechanism: Equilibration of complex numbers}.

\begin{figure}
\includegraphics[width=.8\textwidth]{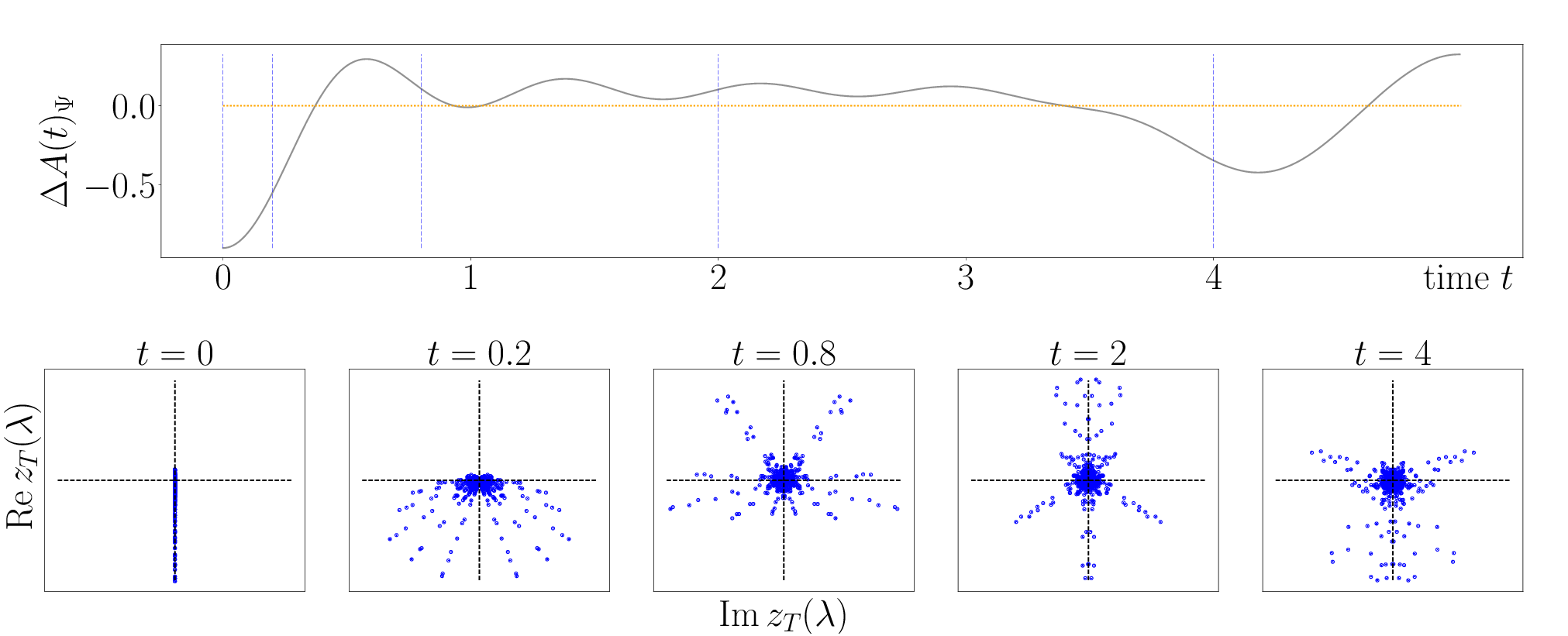}
\caption{Equilibration in the $XX$ chain with its Hamiltonian given in Eq.~\eqref{eq:DefXX} on $L=15$ sites.
We show the exact time
evolution of the deviation of the instantaneous expectation value of a
$\sigma^z$ operator acting on the first site with respect to its steady-state value.
The initial state is a charge density wave state,
i.e., $\ket{1,0,1,0,\ldots,0,1}$, with $\ket{0}$ and $\ket{1}$ denoting the spin up and down state respectively.
In the lower
panel, we plot again the contribution to the Fourier transform of $z_T$ by plotting the evolution of $z_T(\lambda_i)$ in the complex
plane at the times marked in the evolution, where $T\approx 33$ and $\lambda_i$ interpolate the between the larges and smallest gap in $5000$ steps.
Again gaps of the size $|\Delta|<10^{-13}$ are considered to be zero and discarded in order to account for the subtraction of the steady-state value.
}
\label{fig:time_5_ff}
\end{figure}

\begin{figure}
  \includegraphics[width=0.6 \textwidth]{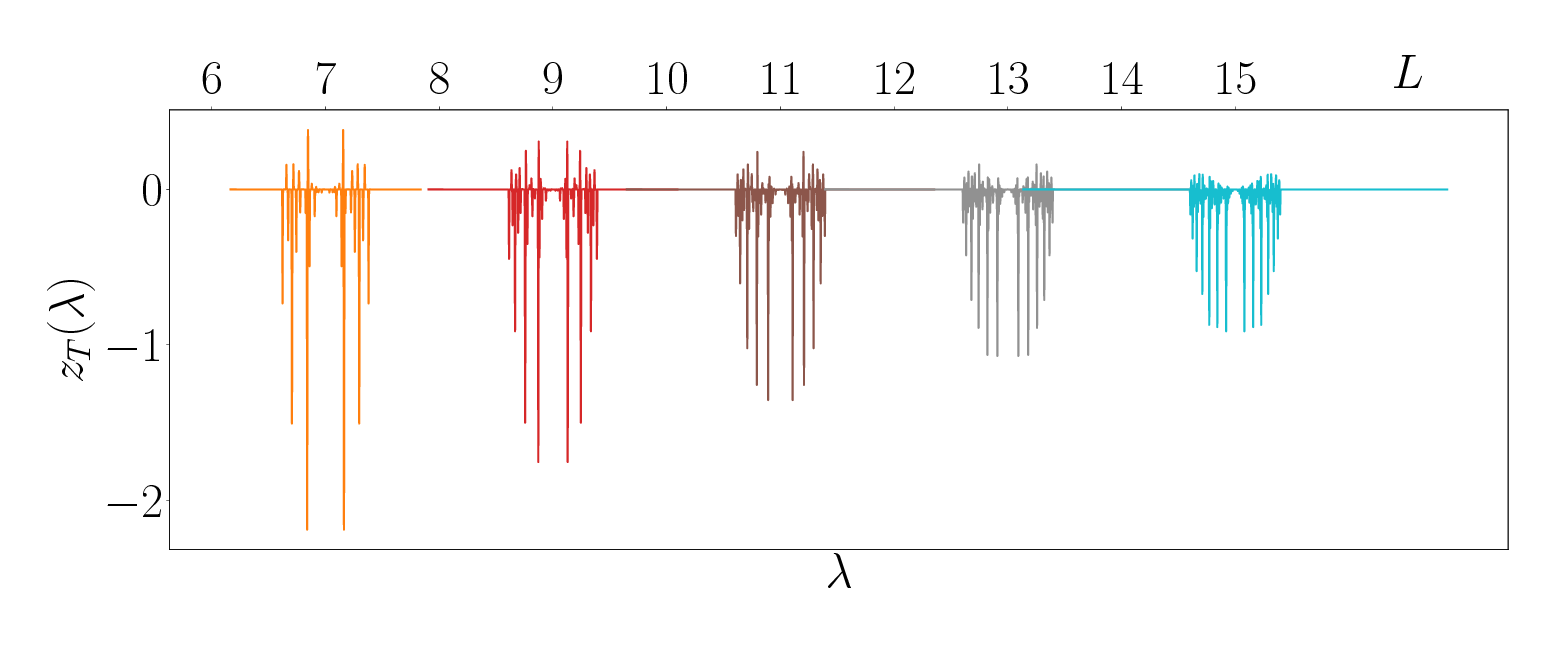}
  \caption{System size scaling of the smoothed $z_T$
  distribution in the $XX$-model with $5000$ points interpolating the minimal and maximal gap with $T\approx33$. Due to the initial
  state being a charge density wave, the results are consistent for odd (or even) system sizes only.}
  \label{fig:G13_ff_var_0_03}
\end{figure}

The first and probably most prominent model is a simple one-dimensional $XX$
spin model
\begin{equation}
  H_\text{XX} = \sum_i \bigg(\sigma^x_i \sigma^x_{i+1}+
  \sigma^y_{i}\sigma^y_{i+1}\bigg),\label{eq:DefXX}
\end{equation}
where $\sigma^x$ and $\sigma^y$ are the spin-$\frac{1}{2}$ Pauli-operators.
The model is exactly solvable by mapping it to a free fermionic hopping model via the
Jordan-Wigner transformation.
This and related models have been studied extensively also in the light of
equilibration (see e.g. \cite{Marek,CalabreseEsslerFagotti11}).
As initial state $\Psi$ we chose a charge density wave-like state with an alternating spin up
and down configuration, i.e., $\ket{1,0,1,0,\ldots,0,1}$, with $\ket{0}$ and $\ket{1}$ denoting the spin up and down state respectively
and as observable $A$ the $\sigma^z$ operator acting on the first site was chosen.
The time evolution of the deviation of the expectation value of $A$ from its infinite time average is shown in Fig.~\ref{fig:time_5_ff}.
We find that the model does equilibrate as expected, however due to
the finite size only up to moderate precision. While theoretically
predicted \cite{CramerEisert}, we cannot confirm a hydrodynamic decay
due to the finite system sizes considered.
The time evolution presented in Fig.~\ref{fig:time_5_ff} suffers from the finite size effect discussed in Sec.~\ref{sec:regularization}.
Namely, $\DelA{t}$ does not drop to zero but quickly decreases and oscillates
around a small but finite value as in the finite system the infinite time average of $\DelA{t}$
does not agree with the infinite time average of $\DelA{t}$ in the infinite system.
In this case, a finite size scaling can consistently be formulated for every second
system size only due to the initial state.
Fig.~\ref{fig:G13_ff_var_0_03} shows the change of the
regularized $z_T$ distributions for the $XX$ model with the system size.
Even for the miniature system size considered we find a visible
shrinking and beginning of smoothening of the $z_T$ distribution indicating
that our main presumption of a bounded, decaying and integrable distribution
appears to be fulfilled better and better for larger system sizes in accordance with the observed equilibration in Fig.~\ref{fig:time_5_ff}.

The second and third model we consider and moreover show in the main text
is the transverse field Ising-Hamiltonian
with local fields taking the form
\begin{equation}
  H_\text{Ising} = \sum_i \bigg(J \sigma^x_i \sigma^x_{i+1} + h_x\sigma^x_{i} +
  h_z\sigma^z_{i}\bigg).\label{eq:AppTransIsing}
\end{equation}
Numerical investigations of the equilibration behavior of this model have been
performed in Refs.\,\cite{Banuls,PollmannTDVP}. While in
Ref.\ \cite{PollmannTDVP} the parameters and initial state is chosen such that
equilibration takes place very rapidly (see Fig.~\ref{fig:EQtime}),
in Ref.\,\cite{Banuls} the authors are
able to find a configuration such that apparently equilibration does not
occur at all (see Fig.~\ref{fig:NEQtime}).
We calculated the $z_\Delta$ distribution using the following
parameters and states. To obtain a rapidly equilibrating system, we use
$J=4,h_x=1,h_z=-2.1$. The initial state is a random product state on all
sites with a single spin in the middle pointing up. The
very slow equilibration is found for the configuration
$J=1,h_x=0.5,h_z=-1.05$ and initial spin-up state on all sites.
In both models we probe as an observable a $\sigma^z$ operator acting on the central site of the chain.
The plots of the corresponding time evolution of the deviation of the
observable from its infinite time average as well as the corresponding
smoothened $z_T$ distributions are shown in the main text in
Fig.~\ref{fig:EQtime}, \ref{fig:NEQtime}, \ref{fig:zdPollmann} and
\ref{fig:zdBanuls} and we discuss their essential features in the captions and the main text.

\begin{figure}
  \includegraphics[width=0.6 \textwidth]{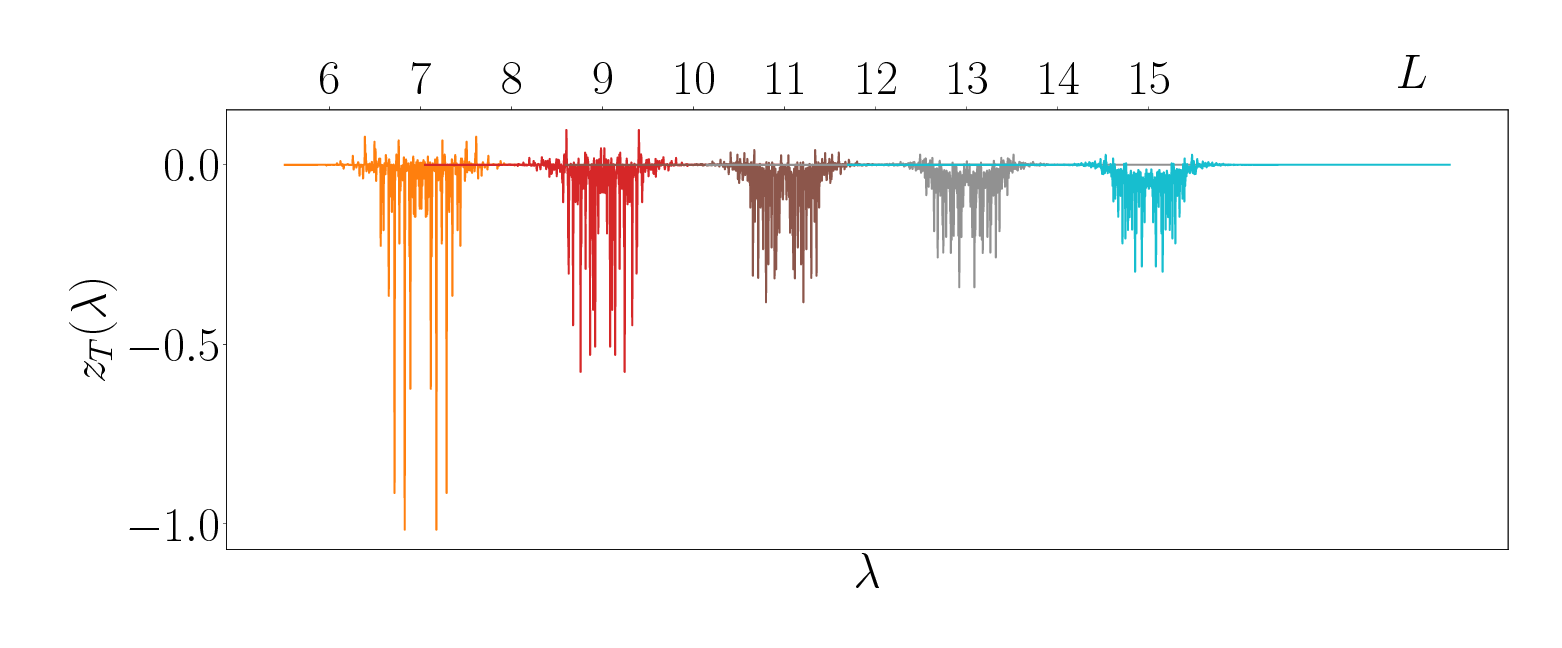}
  \caption{System size scaling of the discretization of the smoothed $z_T$
  distribution in the $XXZ$-model with next-nearest-neighbor hopping when discretized for $5000$ points, where $T\approx33$. Due to the initial
  state being a charge density wave, we only scale the system in
  odd system sizes in order to avoid parity effects.}
  \label{fig:G13_HBnnn_var_0_03}
\end{figure}

The fourth and final model, we considered is the interacting and non-integrable
$XXZ$ or Heisenberg model with next-nearest-neighbor hopping
\begin{equation}
  H_\text{XXZ}  = \sum_i \biggl(J \big(\sigma^x_i \sigma^x_{i+1} +
  \sigma^y_i \sigma^y_{i+1} \big)
  + U \sigma^z_i \sigma^z_{i+1}
  +J_{nnn} \big(\sigma^x_i \sigma^z_{i+1} \sigma^x_{i+2}+ \sigma^y_i
  \sigma^z_{i+1} \sigma^y_{i+2}\big)\biggr),\label{eq:DefHnnn}
\end{equation}
where we use the parameters $J=1,U=2,J_{nnn}=0.2$ and the initial state
is again a charge density wave-like state of alternating spin down and up
configuration and we probe as observable a single $\sigma^z$ operator acting on the first site as in the $XX$ model. 
Let us now consider the same analysis we
have performed for the previous systems.
\begin{figure}
\includegraphics[width=.8\textwidth]{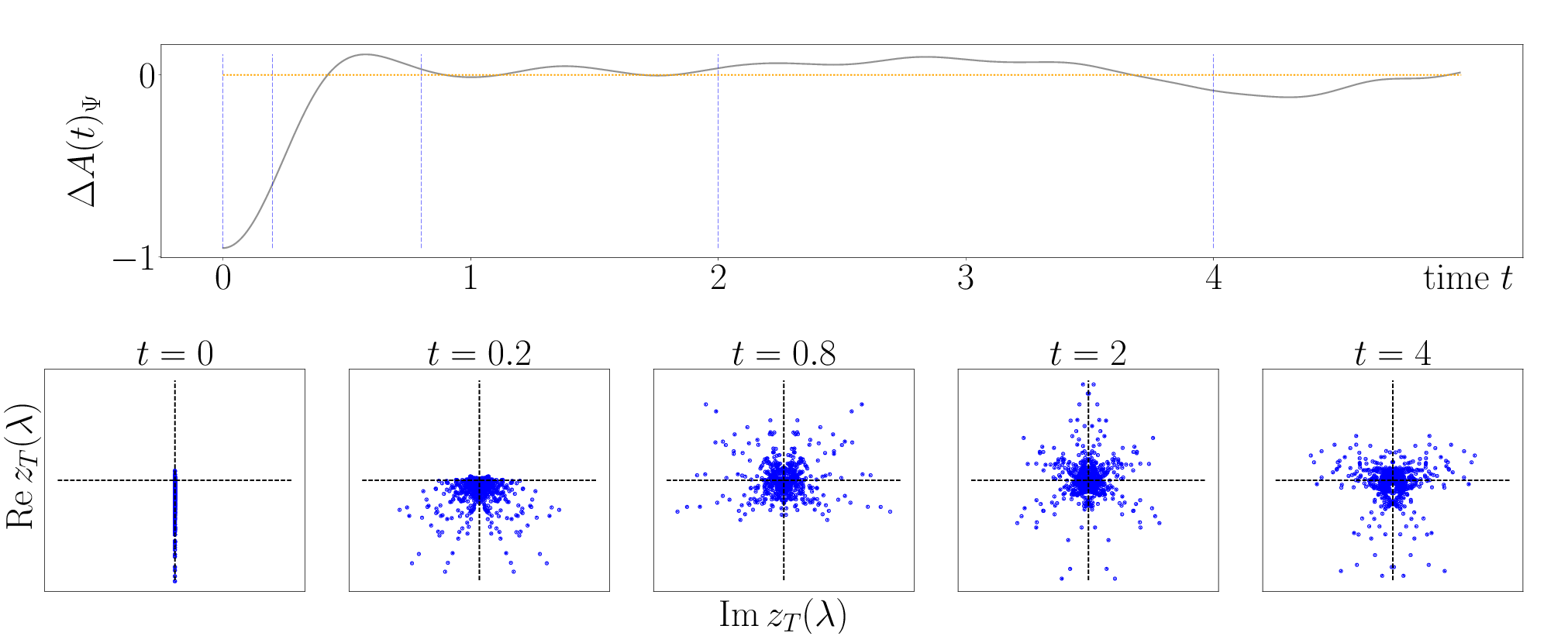}
\caption{Equilibration in the $XXZ$ chain with next-nearest-neighbor hopping defined in Eq.~\eqref{eq:DefHnnn} for $L=15$ sites.
We show the exact time
evolution of the deviation of the instantaneous expectation value of a $\sigma^z$ operator acting on the first site of the chain
with respect to the steady-state value.
The initial state is a charge density wave state.
Moreover, in the lower
panel, we plot the evolution of the $z_T(\lambda_i)e^{\ii t \lambda_i}$ with $\lambda_i$ interpolating the interval between the smallest and largest gap linearly in $5000$ steps and we set $T\approx 33$.
Gaps of the size $|\Delta|<10^{-13}$ are set to zero in order to account for numerical errors when
subtracting the steady-state value.
}
\label{fig:time_5_HBnnn}
\end{figure}
In Fig.~\ref{fig:time_5_HBnnn} we show again the evolution of the deviation of the expectation value of the chosen observable from its infinite time averaged value.
We find for the Heisenberg case a very similar behavior to the $XX$ chain,
although the system is now interacting and non-integrable. The initial
deviation from the steady-state value quickly decays but in contrast
to the non-interacting system, the system size dependent fluctuations are much
weaker.
Similar to the $XX$ chain, the model shows again the finite size effect discussed in Sec.~\ref{sec:regularization} and
$\DelA{t}$ does not drop to zero but due to finite size effects oscillates
around a small but finite value which depends on the system size.
The system size scaling of the $z_T$ distributions displayed in Fig.~\ref{fig:G13_HBnnn_var_0_03}, shows
again a very rapid decrease and smoothening of the distribution again
resembling the $XX$ case.

\begin{figure}
   \includegraphics[width=1\textwidth]{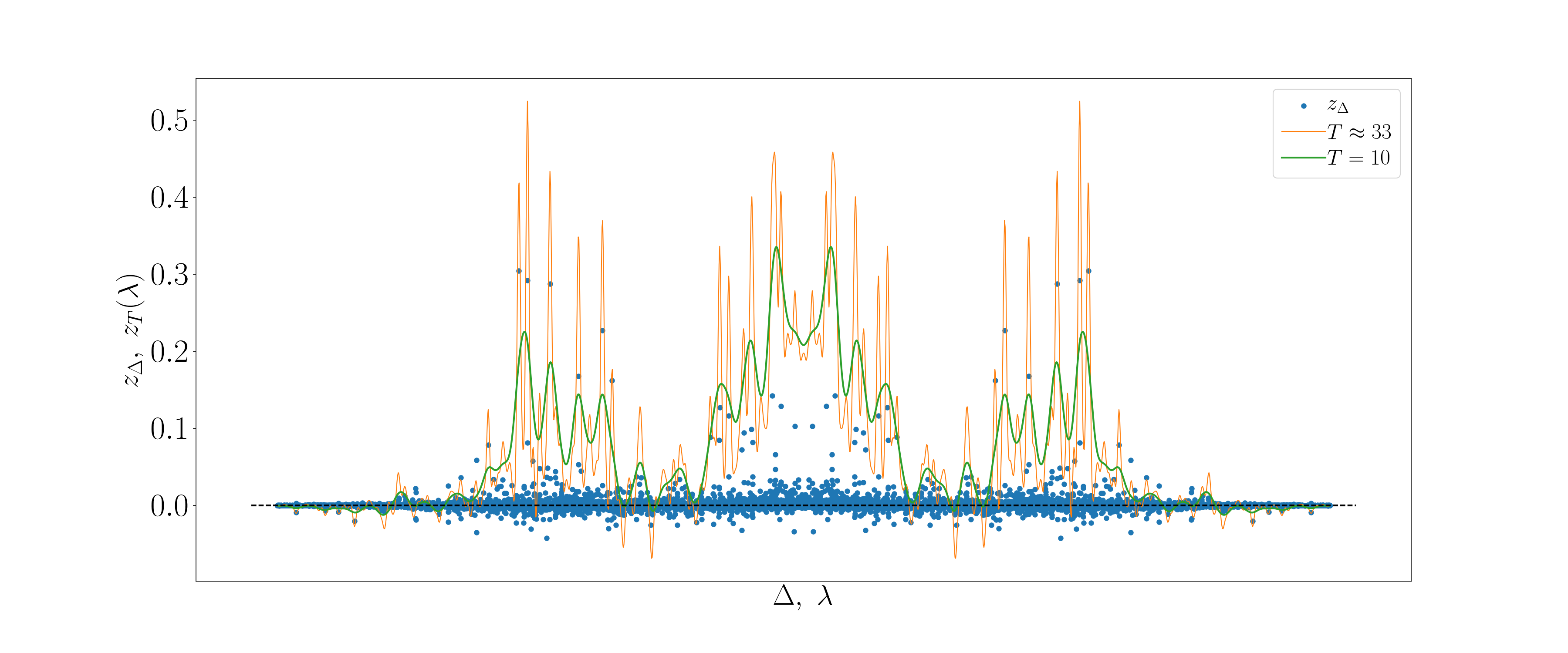}
   \caption{Here, we plot the distribution $z_\Delta$ (discarding $z_0$) and its regularization $z_T(\lambda)$ for $T = 10$ and $T\approx 33$. We again
   discretized $z_T(\lambda)$ using 5000 points between $\Delta_\mathrm{min}$
   and $\Delta_\mathrm{max}$ and considered gaps with $|\Delta|\leq 10^{-13}$
   to be zero. For better visibility, the $z_\Delta$ are scaled by a factor of $10$.}
   \label{fig:complexZDelta}
 \end{figure}

In the last part of this appendix, we would like discuss how the regularization
$z_T(\lambda)$ changes the actual distribution $z_\Delta$ and how severe
finite size effects are for our time evolution.
In Fig.~\ref{fig:complexZDelta} we show the regularized distributions $z_T(\lambda)$ as well as the (scaled)
unregularized data $z_\Delta$ for the equilibrating system presented in Fig.~\ref{fig:EQtime} and \ref{fig:zdPollmann} for system size $10$ and two cut-off times $T=10$ and $T\approx 33$.
By the procedure of regularizing, we reduce the number of points from $2^{2L}$ to $5000$
but more importantly, we keep the relevant features of the distribution intact. This is
to say, that peaks that inhibit equilibration will not be regularized away if they are not accompanied by a respective negative contribution.
As explained in the main text already, the regularization can be interpreted as introducing a cut-off time up to which we expect
equilibration to happen.
For larger cut-off times individual large $z_\Delta$ lead to spikes in the $z_T(\lambda)$ distribution
which are visible in all our presented $z_T(\lambda)$ plots.
In equilibrating systems (cf.~Figs.~\ref{fig:zdPollmann}, \ref{fig:G13_HBnnn_var_0_03} and \ref{fig:G13_ff_var_0_03}) the size and distance of the spikes decreases which leads to a bulk as present in Fig.~\ref{fig:complexZDelta} for small gaps and $\lambda$.

\begin{figure}
  \includegraphics[width=0.75\textwidth]{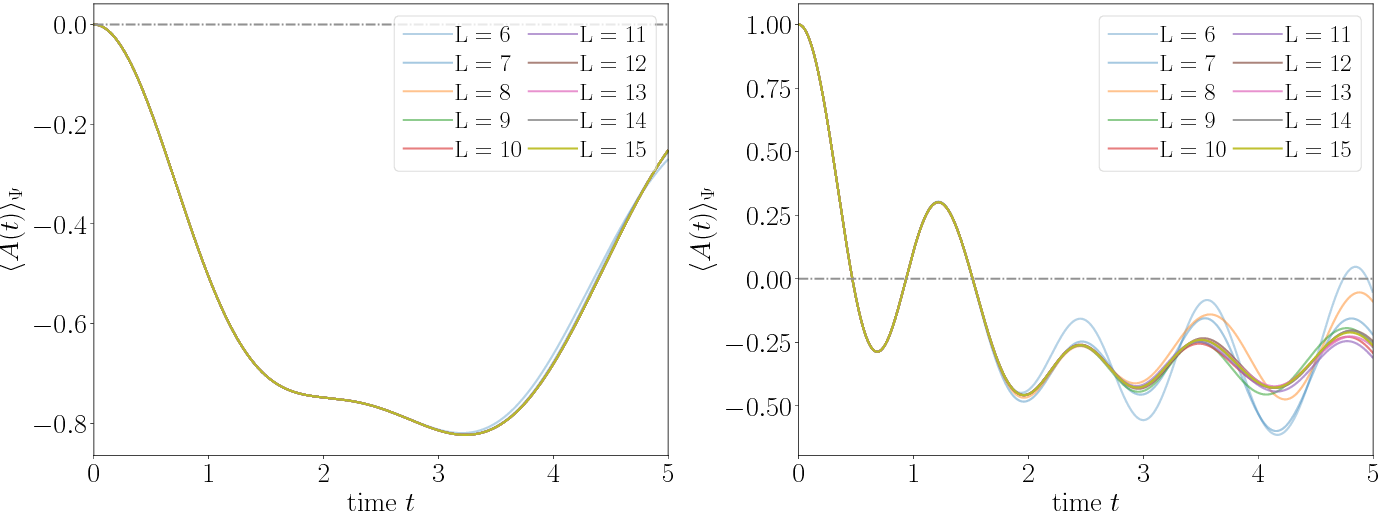}
  \caption{Finite size scaling of the two Ising systems.
  The left plot uses parameters $J=1,h_x=0.5,h_z=-1.05$ and the
  all spin up state, whereas the right plot uses $J=4,h_x=1,h_z=-2.1$
  and a spin up state in the middle surrounded by random product
  states. The observable is in both cases a $\sigma^z_{L/2}$.}
  \label{fig:finTE_Ising}
\end{figure}

\begin{figure}
  \includegraphics[width=0.75 \textwidth]{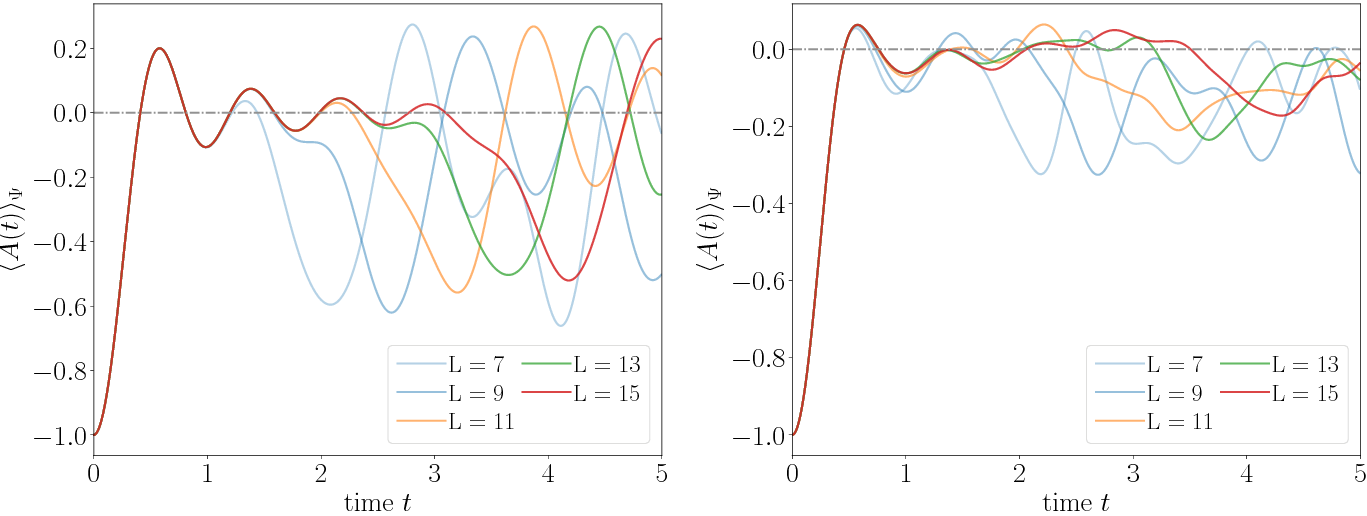}
  \caption{Finite size scaling of the $XX$ (left) and
  the $XXZ$ model (right). Only odd system sizes are considered
  due to the initial state being a charge density wave inducing parity effects between even and odd sites.}
  \label{fig:finTE_XX}
\end{figure}

Finally, let us show and discuss the finite size effects on our time
evolution in the different systems.  In Fig.~\ref{fig:finTE_Ising}, we show the
  evolution of the expectation value $\langle A(t)\rangle_\Psi$
  for different system sizes for the two Ising
configurations from the main text. Fig.~\ref{fig:finTE_XX} shows the same
for the other two considered systems, i.e.~the XX and the Heisenberg model with next-nearest-neighbor hopping.
Note that in contrast to the plots in the main text, we did not subtract the
  steady-state value in these plots. This is due to the size dependence of
  the steady-state value itself that would shift the plots with respect
  to one another. The way we present the system size scaling here instead
focuses on the effect of Lieb-Robinson velocity and emphasizes the time, when
the observable notices the finiteness of the system.

Apart from the non-equilibrating Ising case, all plots show
considerable finite size effects. Due to that and the Lieb-Robinson
velocity, we only plot the time evolution for relatively short times.
However, we also find that as expected the fluctuations
around the equilibrium value decrease with the accessible system sizes.
We would like to stress that we do not extract or use precise data form these plots for our arguments laid out in the main text,
but merely use them to underline our intuition for the essential mechanism of equilibration.
We therefore think that the quite strong finite size effects present are acceptable for our purposes here.

\section{Rigorous results on many-body systems}\label{sec:rigorous}
\subsection{The spectrum}
\label{sec:rigorous:spectrum}
We now turn to collecting rigorous results on quantum lattice models, put together in a notation consistent with the one used in the present work.
Before presenting the precise results, let us first establish the rough intuition underlying the argument.
If we are given a local Hamiltonian $H=\sum_X h_X$, its distribution of energy-eigenstates is simply the probability distribution of energies in the maximally mixed state $\one/d^N$. The maximally mixed state is a product state and the energy distribution is a distribution of a large number of $O(N)$ random variables $h_X$. Since each $h_X$ overlaps only with a finite number of other terms, the random variables can be seen as essentially independent and uncorrelated. It then follows from the central limit theorem that the distribution is Gaussian on large systems, with a standard deviation of order $\sqrt{N}$.
This intuition is made precise in the following Theorem from Ref.\ \cite{BrandaoCramer2015} (for a more direct proof of a similar statement in one-dimensional 
systems, see Ref.\ \cite{Keating2015}).

\begin{theorem}[Berry-Esseen Theorem \cite{BrandaoCramer2015}] Let $H$ be a $k$-local Hamiltonian in $\Lambda=[L]^D$ with $N=L^D$ particles and $\rho$ a state with correlation length $\xi>0$. Let
  \begin{align}
\mu = \tr(\rho H),\quad \sigma = \tr(\rho(H-\mu)^2)^{1/2},\quad s=\frac{\sigma}{\sqrt{N}k^{D/2}}.
  \end{align}
Then
\begin{align}
  \sup_y \left|F(y) - G(y)\leq \Gamma \frac{\log^{2D}(N)}{\sqrt{N}} \right|,
\end{align}
where
\begin{align}
  F(y) := \sum_{k:E_k\leq y}\bra{E_k}\rho\ket{E_k},
  \end{align}
  and
  \begin{align}
    G(y) := \frac{1}{\sqrt{2\pi\sigma^2}}\int_{-\infty}^y \e^{-\frac{(z-\mu)^2}{2\sigma^2}}\,\mathrm{d}z
  \end{align}
  is the Gaussian cumulative distribution with mean $\mu$ and variance $\sigma$ The quantity $\Gamma$ is given by
  \begin{align}
    \Gamma = C_D \frac{(\max\{k,\xi\})^{2D}}{\sigma/\sqrt{N}}\max\left\{\frac{1}{\max\{k,\xi\} \ln(N)},\frac{1}{\sigma^2/N}\right\},
  \end{align}
  where $C_D$ only depends on the dimension of the lattice.
\end{theorem}

Since
\begin{equation}\Gamma \rightarrow C_D \frac{k^{2D}}{\sigma/\sqrt{N}}\max\left\{\frac{1}{k \ln(N)},\frac{1}{\sigma^2/N}\right\}
\end{equation}
 as $\xi\rightarrow 0$, we can now consider a sequence of states $\rho^{(m)}$ that converges to the maximally mixed state and obtain that the distribution of energy-levels, and hence the density of states, converges to a Gaussian in distribution.

%

\subsection{The observable}
\label{sec:rigorous:observable}
We now move to tail bounds for local observables, which show that local observables cannot connect energy eigenstates which differ macroscopically in energy.  This is made precise in the following theorem.

\begin{theorem}[Local observables in energy eigenbasis \cite{LocalGlobal}]
  \label{thm:localglobal}
Let $\Pi_{[\epsilon',\infty)}$ and $\Pi_{[0,\epsilon]}$ be projectors onto the subspaces of energies of $H$ that are
$\geq \epsilon'$ and $\leq \epsilon$, respectively. For a local observable $A$ with $\|A\|=1$, let $H_A$ be the minimal
subset of interaction terms
such that $[H,A] = \sum_{X\in H_A} [h_X, A]$ and let $R:= \sum_{X\in H_A}\|h_X\|$. Then
\begin{eqnarray}
	&& \|  \Pi_{[\epsilon',\infty)} A \Pi_{[0,\epsilon]}\| \leq e^{-\lambda (\epsilon'-\epsilon - 2R)}
	.
	\end{eqnarray}
\end{theorem}
The term $\|A\|$ of the expression in Ref.\ \cite{LocalGlobal}
is not present, as we have normalized the observable.
Note also that in the present work, as in Ref.\ \cite{LocalGlobal}, we assume that the ground state energy is zero.
Theorem~\ref{thm:localglobal} thus shows that matrix-elements of local observables in the energy eigenbasis are exponentially small if the corresponding energies differ by more than an amount which is of the order of the support of the observable.

\subsection{The state}
\label{sec:rigorous:state}
When one quenches from an initial state with short-ranged correlations to a local Hamiltonian, it is expected that the distribution of the overlaps with the eigenstates of the new Hamiltonian will rapidly decay. This is made precise in the following two theorems, first stated for finite correlation lengths and then a stronger theorem for the case of product states. In the following two theorems $n$ denotes the number of terms in the Hamiltonian. Further we denote by $m$ the maximal number of terms $h_{X^\prime}$ with which any $h_X$ does not commute (so shares an overlapping support), i.e.~$m=\max_{X\subset\Lambda}|\{X^\prime:[h_X,h_{X^\prime}]\neq0\}|$.

\begin{theorem}[Finite correlation length \cite{anshuNJP}]
Let $\rho$ be a quantum state with correlation length $\xi>0$ and $\langle H\rangle_\rho=  {\rm tr}(H \rho)$ be the average energy of $\rho$.
For $a\geq (2^{O(D)}/n \xi )^{1/2}$, it holds that
\begin{eqnarray}
	{\rm tr} (\rho \Pi_{[\langle H\rangle_\rho+na,\infty)})
	\leq O(\xi)
	\exp\left(
	-\frac{(n a^2 \xi)^{1/(D+1)}}{O(1) D \xi}
	\right)
\end{eqnarray}
and
\begin{eqnarray}
	{\rm tr} (\rho \Pi_{[0, \langle H\rangle_\rho-na]} ) \leq O(\xi) \exp\left(
	-\frac{(n a^2 \xi)^{1/(D+1)}}{O(1) D \xi}
	\right).
	\end{eqnarray}
\end{theorem}
The theorem hence tells us that in large systems states with a finite correlation length have a sharp distribution in terms of the energy density in the sense that the \emph{total probability} of energies whose energy density differs from the mean energy density decays sub-exponentially.

For product initial states, which take an important role in the discussion of quenches and non-equilibrium physics, the statement is even stronger, giving actual exponential decay:

\begin{theorem}[Product initial states \cite{anshuNJP}]
Let $\rho$ be a product state with  average energy $\langle H\rangle_\rho=  {\rm tr}(H \rho)$.
For $a\geq (O(m^2)/n)^{1/2}$, it holds that
\begin{eqnarray}
	{\rm tr} (\rho \Pi_{[\langle H\rangle_\rho+na,\infty)})
	\leq
	\exp\left(
	-\frac{n a^2}{O(m^2)}
	\right)
\end{eqnarray}
and
\begin{eqnarray}
	{\rm tr} (\rho \Pi_{[0, \langle H\rangle_\rho-na]} ) \leq  \exp\left(
	-\frac{n a^2}{O(m^2)}
	\right),
	\end{eqnarray}
  with $m$ being the maximum numbers of neighbors of any local term in the Hamiltonian.
\end{theorem}

\end{document}